\documentstyle[11pt]{article}
\input epsf

\begin{document}
\def\beq{\begin{equation}}
\def\eeq{\end{equation}}
\vskip30pt

\begin{center}

{\large\bf COMMENTS ON PARTON SATURATION AT SMALL \ x\ IN LARGE NUCLEI }
\vskip20pt
{M.A.Braun\footnote{This research is sponsored in part by the NATO
grant PST.CLG.976799.}\\ Department of High-Energy Physics, S.Petersburg
University\\
198504 S.Petersburg, Russia}
\end{center}

\vskip30pt
\begin{abstract}
It is argued that the gluon density introduced by A.Mueller [5] in
relation to the saturation phenomenon is different from the standardly
defined one, which turned out to have a soliton-like form in the numerical 
calculations. It is shown that A.Mueller's densities obtained from the
latter possess the same saturation properties as originally predicted.
\end{abstract}

\section{Introduction}
\indent 
The idea of parton saturation in large nuclei has recently obtained much
attention [1-5]. In particular A.Mueller has
discussed in [5] the
behaviour of the quark and gluon densities for different values of momenta in
a large nucleus. In his treatment the 
interaction of the probing current and
the nucleons of the target was taken in the Glauber form, 
with the parton-nucleon amplitude taken in the two-gluon exchange
approximation. The evolution of the gluonic distribution in $x$ has not been
studied. The result of these study is
that the behaviour of both the quark and gluonic distribution in the
momentum $l$ is determined by a cerain critical "saturation momentum"
defined as
\beq
Q^2_s=c\frac{8\pi^2\alpha_sN_c}{N_c^2-1}AT(b), 
\eeq
with $T(b)$ the standard nuclear profile function and $c$  essentially
 a constant.
At high momenta  $l^2>>Q_s^2$ the distributions retain their perturbative form
\begin{equation}
{d\big(xq(x,Q^2)\big)
\over d^2bd^2\ell} = {N_c\over 6\pi^4}\  {Q_s^2\over \ell^2}\ 
\end{equation}
and
\begin{equation}
{d\big(xG(x,Q^2)\big)\over d^2bd^2\ell}=
{N_c^2-1\over 4\pi^4}{Q_s^2\over \ell^2}\ln \frac{1}{x}\ 
\ \ \ {\rm for}\  \  \ell^2 >> Q_s^2.
\end{equation}
for the quark and gluon distributions respectively. 
However as the momentum $l$ 
diminishes both distributions do not infinitely grow (as would follow from 
(2) and (3)) 
but saturate at a certain finite value.
In fact at $l^2<<Q_s^2$ A.Mueller found
\begin{equation}
{d\big(xq(x,Q^2)\big)\over d^2bd^2\ell} = {N_c\over 2\pi^4}
\end{equation}
and 
\begin{equation}
{d\big(xG(x,Q^2)\big)\over d^2bd^2\ell}
\approx {N_c^2-1\over 4\pi^4}\ln\frac{1}{x}
\ \ \ {\rm for}\ \  \ell^2 << Q_s^2. 
\end{equation}

Independently of the saturation problem, in the last year 
the  interaction of a probe 
with a large nucleus at
small $x$ was studied in the perturbative QCD with reggeized gluons. 
It reduces to
summing fan diagrams constructed of BFKL pomerons with a 
splitting triple Pomeron
vertex [6-9]. In [9] the resulting equation was solved numerically, 
which enabled us to
calculate the corresponding gluon distribution. Its features turned out to be
somewhat different
from what has been found by A.Mueller. Although at large $l$ the gluonic
distribution indeed falls down similarly  to the perturbative result and to
what has been found in [5], it also falls down to zero at small $l^2$.
The form of the found gluonic distribution as a function of $\log\ l$ is close
to a Gaussian with a constant height and a center moving with a constant 
velocity
to higher $l$ as
the rapidity $y$ increases . Thus in the $(\log\ l,y)$ space the gluonic 
density turns
out to be a soliton wave.

This unexpected behaviour of the gluonic density caused some mistrust due to
the seemingly sound base of the analytic derivation of A.Mueller 
and a purely numerical character of the calculations in [9].
Here we intend to study the reasons for the difference of these results.

One might think that this difference is due to a complicated
gluonic interaction
introduced in the approach pursued in our  numerical calculation
and absent in the picture of A.Mueller.
However the answer seems to be much simpler: the gluonic densities studied in
both papers are in fact different quantities. While the gluonic density
numerically calculated in [9] is more or less standard, the one introduced
by A.Mueller is a new quantity. We show that for this latter density
our numerical results indeed lead to the same saturation as in [5].
 Moreover, A.Mueller's quark density found from our solitonic 
gluon
density also exhibits the same saturation properties
as found in [5].

Besides showing  that there is no contradiction between [5] and [9] at all,
this confirms that the gluonic interaction at small $x$ and
the resulting very sophisticated dynamical picture do not actually 
influence the saturation properties of A.Mueller's densities, as 
conjectured in [5]. 

Of course there remains a question, which is the 
``correct'' parton density. 
As we argue in the last section, we beleive that the densities
introduced by A.Mueller do not in general describe partonic
 distributions in the target.

\section{Gluonic distributions: standard and of A.Mueller}

We begin with the definition of the gluonic density adopted in our numerical
calculations, which we call "standard".
It can be taken from the standard expression
for the structure function of the nucleus in its terms (see e.g [10 ])
\beq
F_2(x,Q^2)=
\frac{4\alpha_sQ^2}{\pi^2N_ce^2}\int d^2b d^2r[\rho_T(r)+\rho_L(r)]
f(x,r,b),
\eeq
where $\rho_{T,L}$ are the transverse and longitudinal colour
densities created by the incoming photon and $f$ is expressed via the
double density of gluons in momentum and
impact parameter:
\beq
f(x,r,b)=\int\frac{d^2k}{(2\pi)^2k^2}
\frac{d \big(xG(x,k^2,b)\big)}{d^2bd k^2}
\left(1-e^{-ikr}\right)\left(1-e^{ikr}\right).
\eeq
In [9] it was found that, up to a factor, fuction $f$ coincides
with a sum of BFKL fan diagrams $\Phi$
\beq
f(x,r,b)=\frac{ N_c}{2\pi^3 \alpha_s}\Phi(x,r,b).
\eeq
Taking a Fourier transform of (7) and neglecting the term proportional to
$\delta^2(k)$ we  obtained
\beq
\frac{d \big(xG(x,k^2,b)\big)}{d^2bd k^2}=
\frac{N_c}{2\pi^2 \alpha_s}k^2\nabla_k^2\phi(x,k,b)
\equiv\frac{ N_c}{2\pi^2 \alpha_s}h(x,k,b),
\eeq
where in the coordinate space
\beq
\phi(x,r,b)=\frac{\Phi(x,r,b)}{2\pi r^2}.
\eeq

Function $\phi$ was found in [9] by numerical solution of
the BFKL fan diagram equation.
Applying $k^2\nabla_k^2$ to it  we found the gluon 
density.  
As mentioned in the Introduction, at any given rapidity the  
 density turned out to have the same, roughly Gaussian,
shape in variable $\xi=\ln k$,
centered at the point $\xi=\xi_0(y)$, which
 moves to the right with a nearly constant velocity. 
Approximately the distribution can be described by 
\beq
h(k)=h_0 e^{-a(\xi-\xi_0(y))^2},
\eeq
where $h_0$ and $a$ are practically independent of $y$ and
$\xi_0(y)$ linearly grows with it:
\beq
h_0\simeq 0.3,\  \ \xi_0(y)=c+ 0.222\ln A+2.23 \frac{\alpha_sN_c}{\pi}y.
\eeq
It can be shown that it is normalized according to
\beq
\int_0^\infty\frac{dk}{k}h(y,k,b)=1,
\eeq
which condition relates $a$ and $h_0$. The starting position of the 
center $c$  depends on the
initial distribution at small rapidities.

Evidently at a given value of $k$ the density always remains limited, 
irrespective of the form of the initial distribution (and on the atomic 
number $A$, in particular). In this sense we have saturation as discussed 
in [1-5]. However with the growth of $y$ the
strongly peaked density moves away toward higher values of $k$ so that the 
density at a fixed point tends to zero at  high values of rapidity.
We thus find  ``supersaturation'': with $y\rightarrow\infty$ the gluon
density
at an arbitrary finite momentum tends to zero.

Now we pass to the gluon density introduced by A.Mueller.
He related it to a particular process, analogous to the $\gamma^*$-target
scattering, in which the electromagnetic current is changed to a 
``gluonic'' current
\begin{equation}
j(x) = - {1\over 4} F_{\mu\nu}^i F_{\mu\nu}^i.
\end{equation}
This current generates a colourless pair of gluons, which according to
[5] interact with the target very much in the same way as a $q\bar q$ pair
 generated by the electromagnetic current 
(see Fig. 1, diagrams $b$ and $c$). The gluon density is then
{\it defined} as a contribution to the cross-section from a given value 
of the transverse momentum $l$ of the  struck gluon 
on the left part of the diagrams.
\begin{figure}
\epsfxsize=15cm \epsfbox{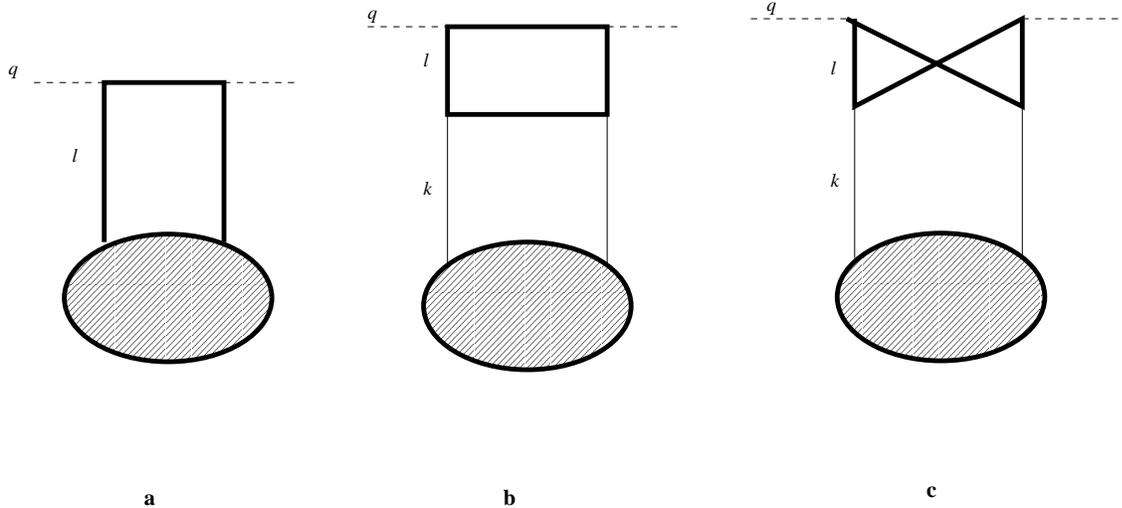}
\caption{Interaction of a probe with the target. Thick lines show partons
(gluons or quarks) directly coupled to the probe. Thin lines show
exchanged gluons.}
\end{figure}
 For the case
when the target is just a nucleon, in the limit $Q^2\rightarrow\infty$
and in the single gluon exchange exchange approximation,
A.Mueller found for his density,
which we denote $xG_M(x,Q^2,l^2)$ (it generally depends on two
different momenta):
\[
{d\big(xG_M(x,Q^2,\ell^2)\big)\over d^2\ell} = {\alpha N_c\over 2\pi^5}
\  \int_{x\ell^2/Q^2}^{\ell^2/Q^2} {dz\over z}\  
{d^2r_1d^2r_2\over r_1^2 r_2^2} \
e^{-i{\ell}\cdot(r_1-r_2)}({2(r_1r_2)^2\over
r_1^2r_2^2}-1)\]\beq 
\cdot  (1+e^{-i{}{k}\cdot({}r_1-{}r_2)}-e^{-i{}{k}\cdot{}r_1}-e^{i{}{k}\cdot{}r_2}){d^2k\over k^2}\ {d \big(xG(x,k^2)\big)\over d k^2}.
\end{equation}
 On the right-hand side
there appeared $ xG(x,k^2)$, which is just the standard gluonic
density. In the approximations made in deriving (15) it does not depend on $x$,
nor the whole expression depends on $Q^2$.
As a result, the gluon density defined by (15)  
also depends on only one momentum squared  $\ell^2$ and superficially
looks similar to the standard gluon density. But in fact it is a different
quantity (factor $\alpha_s$ in front of (15) is already sufficient to
see it). 
Its physical meaning has been explained before:
it is a probability to find one of the struck gluons (the left one in
the Fig. 1)
with momentum $\ell$.
Later we shall discuss to what extent this quantity probes the gluon
distribution in the nucleon.

We see
that once the standard $xG$ is known, the density
$xG_M$ can be found directly from Eq. (15). This exersize is done in the next
section.

\section{The soliton density and saturation}

The first thing to do is to generalize  $xG_M$ to the
nucleus target. Having in mind the fan diagram picture for the interaction
of a probe with the nucleus, on the one hand, and the scattering
process considered by A.Mueller to define his density, on the other,
all we have to do is to change the standard density $xG$ on the right-
hand side of Eq. (15) to the same quantity for the nucleus: 
\begin{displaymath}
{d\big(xG_M(x,Q^2,\ell^2,b)\big)\over{d^2b d^2\ell}} = 
{\alpha N_c\over 2\pi^5}\  \int_{x\ell^2/Q^2}^{\ell^2/Q^2} {dz\over z}\  
{d^2r_1d^2r_2\over r_1^2 r_2^2} \
e^{-i\ell\cdot(r_1-r_2)}({2(r_1r_2)^2\over
r_1^2r_2^2}-1)
 \end{displaymath}
\begin{equation} 
\cdot  (1+e^{-ik\cdot(r_1-r_2)}-
e^{-i{k}\cdot r_1}-e^{i{k}\cdot r_2}){d^2k\over k^2}
\ {d\big(xG\big)\over{d^2b d k^2}}.
\end{equation}
Here we suppresed all evident arguments in $G$ on the right-hand side.
This will be also often  done  in the following when the arguments are
obvious.

We want to calculate this expression taking for
$ d (xG)/d^2b d k^2$ the solitonic 
gluon distribution in a nucleus obtained by our numerical studies.
In principle this can be done rigorously, by numerical integration.
However, since we are interested in the qualitative features of the 
saturation behaviour, we shall adopt here a simpler approach.
We shall use the fact that at large rapidities the solitonic distribution
is strongly peaked at the momena determined by
\beq
k_0=e^{\xi_0}\sim A^{2/9}\exp\left(2.23\frac{N_c\alpha_s}{\pi}y\right)
\eeq
(which are very high already at reasonable $y$ for not very small $\alpha_s$).
This momentum will play the role of Mueller's saturation momentum $Q_s$.
Assuming that $k_0^2<<Q^2$ one 
can take all the integrand out of the integral over $k$
at point $k=k_0$ except for the density factor. The latter, integrated over
$k^2$ gives $N_c/(2\alpha_s\pi^2)$ according to (9) and
the normalization condition (13). So we are
left with the expression
\begin{displaymath}
{d\big(xG_M\big)\over{d^2b d^2\ell}} = 
{N_c^2\over 4\pi^7}\ln\frac{1}{x}  \int d\phi
\  {d^2r_1d^2r_2\over r_1^2 r_2^2} \
e^{-i\ell\cdot(r_1-r_2)}
({2(r_1r_2)^2\over
r_1^2r_2^2}-1)
 \end{displaymath}
\begin{equation} 
\cdot  (1+e^{-i{}{k}\cdot({}r_1-{}r_2)}-
e^{-i{}{k}\cdot{}r_1}-e^{i{}{k},
\cdot{}r_2})
\end{equation}
where one should take $k=k_0n$ with a  two dimensional unit vector $n$.

Calculation of the integrals over $r_1$ and $r_2$ is simplified by using the 
identity
\beq
\frac{1}{r_1^2r_2^2}\left(\frac{2(r_1r_2)^2}
{r_1^2r_2^2}-1\right)=\frac{1}{2}(\nabla_1\nabla_2)^2\ln r_1\ln r_2.
\eeq
One obtains
\beq
{d\big(xG_M\big)\over{d^2b d^2\ell}} = 
{N_c^2\over 8\pi^7}\ln\frac{1}{x}  \int d\phi
\Big[X(l,l)+X(l+k,l+k)-2X(l+k,l)\Big],
\eeq
where
\beq
X(l_1,l_2)=\int d^2r_1d^2r_2 e^{il_1r_1+il_2r_2}
(\nabla_1\nabla_2)^2\ln r_1\ln r_2=4\pi^2\frac{(l_1l_2)^2}{l_1^2l_2^2}.
\eeq
This leads to
\beq
{d\big(xG_M\big)\over{d^2b d^2\ell}} = 
{ N_c^2\over \pi^5}\  \ln\frac{1}{x}
\int d\phi\frac{l^2k_0^2-(lk)^2}{l^2(l+k)^2}.
\eeq 

Let us consider the two limiting cases studied in [5]. If $l>>k_0$ 
then one finds from (22)
\beq
{d(xG_M)\over{d^2b d^2\ell}} = { N_c^2\over \pi^4}\ 
\frac{k_0^2}{l^2} \ln\frac{1}{x}.
\eeq 
In the opposite case $l^2<<k_0^2$ we find
\beq
{d(xG_M)\over{d^2b d^2\ell}} = {N_c^2\over \pi^4}\  \ln\frac{1}{x}
\eeq 
This behaviour is identical with (3) and (5)
(for $N_c>>1$ and
except for factor 4 in (24)).

So Mueller's density  indeed
saturates to a finite value at small $l^2$ irrespective of the
dynamical picture. The standard density however behaves very differently
and for a heavy nucleus goes to zero at any fixed value of $l$ as the
rapidity gets large enough. 

\section{The quark density from the soliton gluonic density}

Following [5] and similarly to the gluon case,  one may introduce a 
quark distribution depending on two momenta squared $xq_M(x,Q^2,\ell^2)$,
 as a contribution from the left struck quark to the
cross-section corresponding to diagrams $b$ and $c$ in Fig. 1.
In the two-gluon exchange approximation for
the interaction, this quantity again results independent of $Q^2$ at
 high virtualities. We postpone  discussion of the relation of this
limiting distribution in $\ell$, $xq_M(x,\ell^2)$ to the standard
quark density as a function of $Q^2$, $xq(x,Q^2)$, until the last section.
Here we shall consider the  properties of $xq_M(x,\ell^2)$ which
follow from the found standard gluon distribution in a large nucleus.

The definition of $xq_M(x,Q^2,\ell^2)$ is very similar to that of (15).
For a single nucleon as a target A.Mueller found:
\beq
{d\big(xq(x,Q^2,\ell^2)\big)\over d^2\ell}  =  {\alpha Q^2\over 4\pi^3}
 \int_0^1 dz [z^2+(1-z)^2]
\left(\frac{\ell^2}{\ell^2+\epsilon^2} -
{(\ell,\ell + k)\over{(\ell^2+\epsilon^2)
[(\ell+k)^2+ \epsilon^2]}}\right){d^2k\over k^2}
\ {d x G\over
d k^2},
\end{equation}
where
\beq
\epsilon^2=Q^2z(1-z).
\eeq
As one observes, the quark density $xq_M$ is again defined via the 
standard gluon density.
 
For the nucleus target, as explained before, all we have to do is to take
the corresponing nuclear density on the right-hand side of (25): 
\beq
{d\big(xq_M\big)\over d^2b d^2\ell}  =  {\alpha Q^2\over 4\pi^3}
 \int_0^1 dz [z^2+(1-z)^2]
\left(\frac{\ell^2}{\ell^2+\epsilon^2} -
{(\ell,\ell + k)\over{(\ell^2+\epsilon^2)
[(\ell+k)^2+ \epsilon^2]}}\right){d^2k\over k^2}
\ {d x G\over
 d^2bd k^2}.
\end{equation}

At high $Q^2$ evidently only small $z$ or $1-z$ contribute. Then
integrating over $\epsilon^2\simeq Q^2z$ we find
\beq
{d\big(xq_M\big)\over d^2b d^2\ell}  =  {\alpha \over 2\pi^3}
\int 
{d^2k\over k^2}
\ {d x G\over
 d^2bd k^2}
\left(1-\frac{(l, l+k)}{(l+k)^2-l^2}\ln\frac{(l+k)^2}{l^2}\right).
\end{equation}

Let us take for $xG$ the
soliton-like gluon density found in [9].
As before we take all smooth functions out of the integral over $k$ and
afterwards integrate over $k$ using (13). We obtain
\begin{equation}
{d\big(xq_M\big)\over {d^2bd^2\ell}} = {N_c \over 4\pi^5} 
\int d\phi
\Big[1-\frac{(l, l+k)}{(l+k)^2-l^2}\ln\frac{(l+k)^2}{l^2}\Big]
\end{equation}  
where $k=k_0n$ as in the preceding section.

Again we shall concentrate on the two limiting cases. If $l>>k_0$
using
\[\frac{1}{(l+k)^2-l^2}\ln\frac{(l+k)^2}{l^2}\simeq
\frac{1}{l^2}\Big(1-\frac{1}{2}\frac{2(lk)+k^2}{l^2}+
\frac{4}{3}\frac{(lk)^2}{l^4}\Big)\]
and integrating over the angle $\phi$ we get
\begin{equation}
{d\big(xq_M\big)\over {d^2bd^2\ell}} = {N_c \over 6\pi^4}
\frac{k_0^2}{l^2}. 
\end{equation}  
In the opposite case $l^2<<k_0$ the second term in the bracket of (29)
does
not contribute and we get
\begin{equation}
{d\big(xq_M\big)\over {d^2bd^2\ell}} = {N_c\over 2\pi^4}.
\end{equation}  
This behaviour exactly coincides with A.Mueller results (2) and (4). 

\section{Discussion}
As we have seen, there are in fact two different gluon densities, which
exhibit different behaviour as functions of momenta. The standard one defined
via Eq (7) has a soliton-like behaviour and vanishes at any given $\ell$
as $y\rightarrow\infty$. The other, defined in [5], 
saturates at small momenta to
 a finite value, 
which turns out to be universal in the sense that it does not depend
on the complexity of the interaction mechanism. Which of the two
densities is the ``correct'' one?
   
The definition of the standard gluon density via Eq. (7) seems to us
perfectly adequate. Its validity is clear from the inspection of the diagrams
$b$ and $c$ of the Fig. 1 corresponding to (7). The lower gluonic blob, to
which the quark loop is coupled, has to be integrated over the ``-'' component
 of the 4-momentum k (in a system, in which the target has momentum with
a large ``+'' component). Evidently ihe blob then represents just the gluonic
density at a given ``time'' $x_+$.

Interpretation of Mueller's density, in contrast, is not so obvious.
First, in the presence of the interference diagram $c$ of Fig. 1,
one may hardly say that the contribution from the left struck gluon
to the cross-section probes the gluon distribution in the target. In the
interference diagram the right struck gluon has a different transverse
momentum $\ell+k$, so that the interpretation of the whole contribution as
a gluon density is doubtful. Then it is not clear why the lowest order 
diagram Fig. 1$a$ has not been included. Finally, in the BFKL kinematics
the virtual gluons tend to reggeize. One then expects that the interference
digram can be neglected and diagram Fig. 1$b$ will be absorbed into $a$
to form  a BFKL pomeron coupled to the target. So in the end one will find 
only the diagram Fig. 1$a$, which leads to the 
standard gluon density as defined through 
Eq. (7).

The same  doubts refer to the quark density in the definition of A.Mueller.
Again, so long as one cannot neglect the interference diagram, the
interpretation of the contribution to the cross-section
from the left struck quark with
a given momentum as a quark density of the target does not seem possible.
True, in the DGLAP kinematics, with finite $x$ , the interference
diagram can indeed be neglected, substituted by the requirement that
$\ell^2<<Q^2$. However then one has the transverse momentum ordering
and the density found from Eq. (25) would be trivial
(see [5], Eq. (14)). In the BFKL regime, at small $x$,
the virtual quarks also tend to reggeize and the contribution from
digram Fig. 1$c$ hopefully becomes neglegible as compared to $a$ and $b$,
which actually become identical. Then the
interpretation of (25) as the quark density may aquire sense.
However to calculate it one should couple the found gluonic distribution
to a reggeized quark. We leave this problem for future studies.

\section{References}

\hspace*{0.6 cm}1. L.V. Gribov, E.M. Levin and 
M.G. Ryskin, Phys.Rep.{\bf
100} (1983)1.

2. A.H. Mueller, Nucl.Phys.{\bf B335} (1990) 115.

3. J.Jalilian-Marian, A. Kovner, L.McLerran and H. Weigert, 
Phys.Rev.D{\bf 55} (1997) 5414.

4. Yu. V. Kovchegov and A.H. Mueller, Nucl.Phys.{\bf B529} (1998) 451.

5. A.H.Mueller, hep-ph/9904404

6. I.Balitsky, hep-ph/9706411; Nucl. Phys. {\bf B463} (1996) 99.

7. Yu. Kovchegov, Phys. Rev {\bf D60} (1999) 034008;\\
  preprint CERN-TH/99-166 (hep-ph/9905214).

8.E.Levin and K.Tuchin, preprint DESY 99-108, TAUP 2592-99
(hep-ph/9908317).
 
9. M.A.Braun, Eur.Phys. J. {\bf C16} (2000) 337-348.

10. M.A.Braun, Eur. Phys. J {\bf C6} (1999) 343.\\
\end{document}